\documentclass[12pt]{article}

\usepackage{amsmath,amsthm,amssymb,amsfonts}
\usepackage{mathrsfs}
\usepackage[latin1]{inputenc}
\usepackage{listings}
\usepackage[dvips]{color}
\usepackage{xcolor}
\usepackage[ruled]{algorithm2e} 
\usepackage{calc}
\usepackage[a4paper,left=2.5cm, right=2.5cm, bottom=2.5cm, top=2.5cm]{geometry}
\usepackage{appendix}
\usepackage{rotating}
\usepackage{subfigure}
\usepackage{epsfig}
\usepackage{natbib}
\usepackage{url}
\usepackage{verbatim}
\usepackage[section] {placeins}
\usepackage[symbol*]{footmisc}
\usepackage{pifont}
\usepackage{setspace}
\usepackage{lscape}
\usepackage{float}
\usepackage{color}
\usepackage{longtable}

\newtheorem{theorem}{Theorem}

\newcommand{\blind}{0}

\def\signed #1{{\leavevmode\unskip\nobreak\hfil\penalty50\hskip2em
  \hbox{}\nobreak\hfil(#1)%
  \parfillskip=0pt \finalhyphendemerits=0 \endgraf}}

\newsavebox\mybox

\newcommand{\Xb}{\boldsymbol{X}}
\newcommand{\xb}{\boldsymbol{x}}
\newcommand{\hm}{\hat{\boldsymbol{\mu}}}
\newcommand{\hS}{\hat{\boldsymbol{\Sigma}}}

\newcommand{\argmin}{\mathop{\mbox{argmin}}}

\newcommand{\Db}{\boldsymbol{D}}
\newcommand{\Eb}{\boldsymbol{E}}

\newcommand{\Vb}{\boldsymbol{V}}

\newcommand{\Sb}{\boldsymbol{S}}

\newcommand{\Ub}{\boldsymbol{U}}
\newcommand{\Wb}{\boldsymbol{W}}
\newcommand{\Yb}{\boldsymbol{Y}}
\newcommand{\Zb}{\boldsymbol{Z}}

\newcommand{\zb}{\boldsymbol{z}}
\newcommand{\yb}{\boldsymbol{y}}
\newcommand{\ub}{\boldsymbol{u}}

\newcommand{\wb}{\boldsymbol{w}}
\newcommand{\mb}{\boldsymbol{m}}

\newcommand{\hmrewo}{\hat{\boldsymbol{\mu}}_\textrm{OGK}}
\newcommand{\hSrewo}{\hat{\boldsymbol{\Sigma}}_\textrm{OGK}}

\begin{document}

\def\spacingset#1{\renewcommand{\baselinestretch}%
{#1}\small\normalsize} \spacingset{1}


\if0\blind
{
  \title{\bf The Minimum Regularized Covariance Determinant estimator}
  \author{Kris Boudt\thanks{
    This research has benefited from the financial support of the Flemish Science Foundation (FWO) and project C16/15/068 of Internal Funds KU Leuven.
We are grateful to {Valentin Todorov for adding the MRCD functionality to the R package {\it rrcov} \citep{rrcov}, and to} Yukai Yang for his initial assistance to this work. We also thank Dries Cornilly, Christophe Croux,  Gentiane Haesbrouck, Sebastiaan H\"oppner, Stefan Van Aelst and Marjan Wauters for their constructive comments.}\hspace{.2cm}\\
    Solvay Business School, Vrije Universiteit Brussel\\
		School of Business and Economics,  Vrije Universiteit Amsterdam\\
    \\
    Peter J. Rousseeuw  \\
    Department of Mathematics, KU Leuven\\
		\\
		Steven Vanduffel \\
		Solvay Business School, Vrije Universiteit Brussel\\
    \\
		Tim Verdonck\\
		Department of Mathematics, KU Leuven\\
		}
	\date{November 29, 2018}
  \maketitle
} \fi

\if1\blind
{
  \bigskip
  \bigskip
  \bigskip
  \begin{center}
    {\LARGE\bf The Minimum Regularized Covariance Determinant estimator}
\end{center}
  \medskip
} \fi

\bigskip
\begin{abstract}
The Minimum Covariance Determinant (MCD) approach estimates the location and scatter matrix using the subset of given size with lowest sample covariance determinant. Its main drawback is that it cannot be applied when the dimension exceeds the subset size. We propose the Minimum Regularized Covariance Determinant (MRCD) approach, which differs from the MCD in that the { scatter} matrix is a convex combination of a target matrix and the sample covariance matrix { of the subset}. A data-driven procedure sets the weight of the target matrix, so that the regularization is only used when needed. The MRCD estimator is defined in any dimension, is well-conditioned by construction and preserves the good robustness properties of the MCD. We prove that so-called concentration steps can be performed to reduce the MRCD objective function, and we exploit this fact to construct a fast algorithm. We verify the accuracy and robustness of the MRCD estimator in a simulation study and illustrate its practical use for outlier detection and regression analysis on real-life high-dimensional data sets in chemistry and criminology.
\end{abstract}

\noindent%
\vfill

\newpage
\spacingset{1.45} 

\section{Introduction}
The Minimum Covariance Determinant (MCD)
method~\citep{rousseeuw1984least,rousseeuw1985multivariate}
is a highly robust estimator of multivariate location and
scatter.
Given an $n\times p$ data matrix
$\Xb=(\xb_1,\ldots,\xb_n)'$ with $\xb_i=(x_{i1},\ldots,x_{ip})'$,
its objective is to find $h$ observations
whose sample covariance matrix has the lowest possible determinant.
Here $h < n$ is fixed.
The MCD estimate of location is then the average of these $h$ points,
whereas the scatter estimate is a multiple of their covariance matrix.
Consistency and asymptotic normality of the MCD estimator have
been shown
by~\citet{Butler:AsymptMCD} and~\citet{Cator:AsympMCD}.
The MCD has a bounded influence function~\citep{crouxhaes1999}
and has the highest possible breakdown value (i.e.\ $50\%$) when
$h=\left\lfloor (n+p+1)/2 \right\rfloor$~\citep{Lopuhaa:BDP}.
The MCD approach has been applied to various fields such as
chemistry, finance, image analysis, medicine, and quality control,
see e.g. the review paper of \cite{Hubert:ReviewHighBreakdown}.

A major restriction of the MCD approach is that the dimension
$p$ must satisfy $p<h$ for the covariance matrix of any
$h$-subset to be non-singular.
In fact, for accuracy of the estimator it is often recommended
to take $n > 5p$, e.g. in \citet{rousseeuw2015robustbase}.
This limitation creates a gap in the availability of high
breakdown methods for so-called ``fat data'', in which the
number of rows (observations) is small compared to the number
of columns (variables).
To fill this gap we propose a modification of the MCD to make
it applicable to high dimensions. The basic idea is to replace
the subset-based covariance by a regularized covariance estimate, defined as a weighted average of the sample covariance of the
$h$-subset and a predetermined positive definite target matrix.
The proposed Minimum Regularized Covariance Determinant (MRCD) estimator is then the regularized covariance based on the $h$-subset which makes the overall determinant the smallest.

In addition to its availability for high dimensions, the main features of the MRCD estimator are that it preserves the good
breakdown properties of the MCD estimator and is
well-conditioned by construction.
Since the estimated covariance matrix is guaranteed to be
invertible it is suitable for computing robust distances, and
for linear discriminant analysis and graphical
modeling \citep{ollerer2015robust}.
Furthermore, we will generalize the C-step theorem
of \citet{RD:99} by showing that the objective function is
reduced when concentrating the $h$-subset to the $h$
observations with the smallest robust distance computed from
the regularized covariance.
This C-step theorem forms the theoretical basis for the proposed
fast MRCD estimation algorithm.

The remainder of the paper is organized as follows.
In Section \ref{MCDtoMRCD} we introduce the MRCD covariance estimator and discuss its properties.
Section \ref{Estimation} proposes a practical and fast algorithm
for the MRCD. The extensive simulation study in Section
\ref{Simulation} confirms the good properties of the method.
Section \ref{Applications} uses the MRCD estimator for outlier detection and regression analysis on real data sets from
chemistry and criminology. The main findings and suggestions for further research are summarized in the conclusion.

\section{From MCD to MRCD}
\label{MCDtoMRCD}

Let $\xb_1,\ldots,\xb_n$ be a dataset in which
$\xb_i=(x_{i1},\ldots,x_{ip})'$ denotes the $i$-th observation
($i=1,\ldots,n$). The observations are stored in the $n\times p$ matrix $\Xb=(\xb_1,\ldots,\xb_n)'$. We assume that most of them
come from an elliptical distribution with location
$\boldsymbol{\mu}$ and scatter matrix $\boldsymbol{\Sigma}$.
The remaining observations can be arbitrary outliers,
and we do not know beforehand which ones they are.
The problem is to estimate $\boldsymbol{\mu}$
and $\boldsymbol{\Sigma}$ despite the outliers.

\subsection{The MCD estimator}

The MCD approach searches for an $h$-subset of the data
(where $n/2 \leqslant h < n$) whose sample covariance matrix
has the lowest possible determinant.
Clearly, the subset size $h$ affects the efficiency of the estimator as well as its robustness to outliers.
For robustness, $n-h$ should be at least the number of outliers.
When many outliers could occur one may set $h=\lceil 0.5n\rceil$.
Typically one sets $h=\lceil 0.75n\rceil$ to get a better
efficiency.
Throughout the paper, $H$ denotes a set of $h$
indices reflecting the observations included in the subset,
and $\mathcal{H}_h$ is the collection of all such sets.
For a given $H$ in $\mathcal{H}_h$ we denote the corresponding
$h\times p$ submatrix of $\mathbf{X}$ by $\mathbf{X}_H$.  Throughout the paper, we use the term $h$-subset to denote both $H$ and $\mathbf{X}_H$ interchangeably.  The mean and sample covariance matrix of $\mathbf{X}_H$ are
then
\begin{eqnarray}
\mathbf{m}_{\Xb}(H) &=& h^{-1}
     \mathbf{X}_H' \mathbf{1}_h
		 \label{meanh} \\
\mathbf{S}_{\Xb}(H) &=& (h-1)^{-1}
    (\mathbf{X}_H - \mathbf{m}_{\Xb}(H))'
    (\mathbf{X}_H - \mathbf{m}_{\Xb}(H))\;\;.
		\label{covarianceh}
\end{eqnarray}
The MCD approach then aims to minimize the determinant of
$\mathbf{S}_{\Xb}(H)$ among all $H \in \mathcal{H}_h$:
\begin{equation}
H_{MCD}= \argmin_{H \in \mathcal{H}_h}
  \left(\det(\mathbf{S}_{\Xb}(H))^{1/p}\right) \label{subset1star}
\end{equation}
where we take the $p$-th root of the determinant for
numerical reasons. Note that the $p$-th root of the
determinant of the covariance matrix is the geometric mean
of its eigenvalues; \citet{sengupta1987tests} calls it the
standardized generalized variance.
The MCD can also be seen as a multivariate least trimmed
squares estimator in which the trimmed observations have the largest Mahalanobis distance with respect to
the sample mean and covariance of the $h$-subset
\citep{ACA:08}.

The MCD estimate of location $\mathbf{m}_{MCD}$ is defined
as the average of the $h$-subset, whereas the MCD scatter
estimate is given as a multiple of its sample covariance
matrix:
\begin{eqnarray}
  \mathbf{m}_{MCD} &=& \mathbf{m}_{\Xb}(H_{MCD})
	\label{eq:MCD1}\\
\mathbf{S}_{MCD} &=& c_\alpha \mathbf{S}_{\Xb}(H_{MCD})
  \label{eq:MCD2} \label{eq:cfactorMCD}
\end{eqnarray}
where $c_\alpha$ is a consistency factor
such as the one given by \citet{crouxhaes1999},
and depends on the trimming percentage $\alpha = (n-h)/n$.
\citet{Butler:AsymptMCD} and~\citet{Cator:AsympMCD} prove
consistency and asymptotic normality of the MCD estimator,
and \citet{Lopuhaa:BDP} show that it has the highest possible
breakdown value (i.e.,\ $50\%$) when
$h=\left\lfloor (n+p+1)/2 \right\rfloor$.
Accurately estimating a covariance matrix requires a
sufficiently high number of observations.
A rule of thumb is to require $n>5p$ \citep{rousseeuw1990unmasking,rousseeuw2015robustbase}.
When $p>h$ the MCD is ill-defined since all
$\mathbf{S}_{\Xb}(H)$ have zero determinant.

\subsection{The MRCD estimator}

We will generalize the MCD estimator to high dimensions.
As is common in the literature, we first
standardize the $p$ variables { to ensure that the final MRCD scatter estimator is location invariant and scale equivariant. This means that for any diagonal $p\times p$
matrix $\mathbf{A}$ and any $p\times 1$ vector $\mathbf{b}$
the MRCD scatter estimate $S(\mathbf{A} \mathbf{X} + \mathbf{b})$
equals $\mathbf{A}\mathbf{S}(\mathbf{X})\mathbf{A}'$\;. The standardization needs to use  a robust univariate location and scale estimate. To achieve this,}  we compute the median of each variable and stack
them in a location vector $\mathbf{\nu}_{\Xb}$.
We also estimate the scale of each variable by the Qn
estimator
of \citet{rousseeuw1993alternatives}, and put these scales
in a diagonal matrix $\mathbf{D}_{\Xb}$.
The standardized observations are then
\begin{equation}
  \ub_i = \mathbf{D}^{-1}_{\Xb}(\xb_i-\mathbf{\nu}_{\Xb})\;\;.
	\label{uobs}
\end{equation}
This disentangles the location-scale and
correlation problems, as in
\citet{boudt2012jump}. 

In a second step, we use a predetermined and well-conditioned
symmetric and positive definite target matrix $\mathbf{T}$.
We also use a scalar weight coefficient $\rho$, henceforth
called the regularization parameter.
We then define the regularized covariance matrix of
an $h$-subset $H$  of the standardized data $\Ub$  as
\begin{equation}
  {\mathbf{K}}(H) = \rho \ \mathbf{T} + (1-\rho)
  c_\alpha \mathbf{S}_{\Ub}(H)
	\label{subshrinkage1}
\end{equation}
where $\mathbf{S}_U(H)$ is as defined in (\ref{covarianceh})
but for $\Ub$,
and $c_\alpha$ is the same consistency factor as in
(\ref{eq:cfactorMCD}).

It will be convenient to use the { spectral} decomposition
$\mathbf{T}=\mathbf{Q}\mathbf{\Lambda}\mathbf{Q}'$
where $\mathbf{\Lambda}$ is the diagonal matrix holding the
eigenvalues of $\mathbf{T}$ and $\mathbf{Q}$ is the orthogonal
matrix holding the corresponding eigenvectors.
We can then rewrite the regularized covariance matrix
$\mathbf{K}(H)$ as
\begin{equation}
  {\mathbf{K}}(H) = \mathbf{Q}\mathbf{\Lambda}^{1/2}
  [\rho \ \mathbf{I} + (1-\rho) c_\alpha \mathbf{S}_{\Wb}(H)]
	\mathbf{\Lambda}^{1/2} \mathbf{Q}'
	\label{subshrinkage}
\end{equation}
where the $n \times p$ matrix $\Wb$ consists of the
transformed standardized observations
$\wb_i = \mathbf{\Lambda}^{-1/2}\mathbf{Q}'\ub_i.$
It follows that
$\mathbf{S}_{\Wb}(H) = \mathbf{\Lambda}^{-1/2}\mathbf{Q}'
 \mathbf{S}_{\Ub}(H)\mathbf{Q} \mathbf{\Lambda}^{-1/2}$.

The MRCD subset $H_{MRCD}$ is defined by minimizing the
determinant of the regularized covariance matrix
$\mathbf{K}(H)$ in (\ref{subshrinkage}):
\begin{equation}
  H_{MRCD}= \argmin_{H \in \mathcal{H}_h}
	          \left(\det(\mathbf{K}(H))^{1/p}\right)\;\;.
	\label{subset2star}
\end{equation}
Since $\mathbf{T}$, $\mathbf{Q}$ and $\mathbf{\Lambda}$ are
fixed, $H_{MRCD}$ can also be written as
\begin{equation}
  H_{MRCD}= \argmin_{H \in \mathcal{H}_h}
	 \left(\det(\rho \ \mathbf{I} + (1-\rho)
	c_\alpha \mathbf{S}_{\Wb}(H))^{1/p}\right)\;\;.
\label{subset2star1}
\end{equation}
Once $H_{MRCD}$ is determined, the MRCD location and scatter estimates of the
original data matrix $\mathbf{X}$ are defined as
\begin{eqnarray}
  \mathbf{m}_{MRCD}  &=& \mathbf{\nu}_{\Xb}+
	    \mathbf{D}_{\Xb}\mathbf{m}_{\Ub}(H_{MRCD}) \\
  \mathbf{K}_{MRCD}  &=& \mathbf{D}_{\Xb} \mathbf{Q}
	    \mathbf{\Lambda}^{1/2}
			[\rho \ \mathbf{I} + (1-\rho)\mathbf{S}_{\Wb}c_{\alpha}(H_{MRCD}) ]
			\mathbf{\Lambda}^{1/2} \mathbf{Q}' \mathbf{D}_{\Xb}.
  \label{MRCDestimates}
\end{eqnarray}

{The MRCD is not affine equivariant, as this would require that $S(\mathbf{A} \mathbf{X} + \mathbf{b})$ equals $\mathbf{A}\mathbf{S}(\mathbf{X})\mathbf{A}'$ for all nonsingular matrices $A$ and any $p\times 1$ vector $\mathbf{b}$. As mentioned before,
the MRCD scatter estimate is location invariant and scale equivariant due to the initial standardization step }

\subsection{The MRCD precision matrix }\label{subsec:inverse}

The precision matrix is the inverse of the scatter matrix, and
is needed for the calculation of robust MRCD-based Mahalanobis
distances, for linear discriminant analysis, for graphical
modeling \citep{ollerer2015robust}, and for many other
computations.
 {  By (\ref{MRCDestimates}) the MRCD precision matrix is given by the expression}
\begin{eqnarray}
  \mathbf{K}_{MRCD}^{-1} &=& \mathbf{D}_{\Xb}^{-1}\mathbf{Q}'
	\mathbf{\Lambda}^{-1/2}
	[\rho \ \mathbf{I}_p + (1-\rho)c_{\alpha}\mathbf{S}_{\Wb}(H_{MRCD})]^{-1}
	\mathbf{\Lambda}^{-1/2} \mathbf{Q}\mathbf{D}_{\Xb}^{-1}\;\;.
  \label{invMRCDestimates}
\end{eqnarray}
{ When $p>h$, a computationally more convenient form can be obtained by the Sherman-Morrison-Woodbury identity \citep{sherman1950adjustment,woodbury1950inverting,bartlett1951inverse} as follows:
\begin{eqnarray}
 \mathbf{K}_{MRCD}^{-1}
  &=& \mathbf{D}_{\Xb}^{-1}\mathbf{Q}'
  \mathbf{\Lambda}^{-1/2}
 \left[\frac{1}{\rho}\mathbf{I}_p-\frac{1}{\rho^2}
 \frac{(1-\rho)c_{\alpha}}{h-1}\boldsymbol{Z}'\left(\mathbf{I}_h+\frac{1}{\rho}\frac{(1-\rho)c_{\alpha}}{h-1}
 \boldsymbol{Z}\boldsymbol{Z}'\right)^{-1}
 \boldsymbol{Z}\right] \mathbf{\Lambda}^{-1/2} \mathbf{Q}\mathbf{D}_{\Xb}^{-1}
\label{invMRCDestimates2}
\end{eqnarray}
where $Z=\Wb_{H_{MRCD}}-\mathbf{m}_{\Wb}(H_{MRCD})$ and hence } {$\mathbf{S}_{\Wb}(H_{MRCD})=\Zb'\Zb/(h-1)$. Note that the advantage of (\ref{invMRCDestimates2}) is that only a $h\times h$ matrix needs to be inverted, rather than
a $p\times p$ matrix as in (\ref{invMRCDestimates}). }

The MRCD should not be confused with the
Regularized Minimum Covariance Determinant (RMCD)
estimator of \citet{croux2012}.
The latter assumes sparsity of the precision matrix, and
maximizes the penalized log-likelihood function of each
$h-$subset by the GLASSO algorithm of \citet{glasso2008}.
The repeated application of GLASSO is time-consuming.

\subsection{Choice of target matrix and calibration
            of $\rho$}\label{subsec:target}

The MRCD estimate depends on two quantities: the target
matrix $\mathbf{T}$ and the regularization parameter $\rho$.
For the target matrix $\mathbf{T}$ on $\Ub$ we can take
the identity matrix; relative to the original data $\Xb$
this is the diagonal matrix with the robustly estimated
univariate scales on the diagonal.
Depending on the application, we can also take a non-diagonal target matrix $\mathbf{T}$.
When this matrix is estimated in a first step, it should be
robust to outliers in the data.
A reasonable choice is to compute a rank correlation
matrix of $\Ub$, which incorporates some of the relation
between the variables.
When we have reasons to suspect an equicorrelation
structure, we can set $\mathbf{T}$ equal to
\begin{equation}
  \mathbf{R}_c =c \mathbf{J}_p+(1-c)\mathbf{I}_p
\label{equicor}
\end{equation}
with $\mathbf{J}_p$ the $p\times p$ matrix of ones,
$\mathbf{I}_p$ the identity matrix,
and $-1/(p-1)< c < 1$ to ensure positive definiteness.
The parameter $c$ in the equicorrelation matrix
(\ref{equicor}) can be estimated by averaging robust
correlation estimates over all pairs of variables,
under the constraint that the determinant of $\mathbf{R}_c$
is above a minimum threshold value.

When the regularization parameter $\rho$ equals zero
$\mathbf{K}(H)$ becomes the sample covariance
$\mathbf{S}_{\Ub}(H)\;$, and when $\rho$ equals one
$\mathbf{K}(H)$ becomes the target.
We require  $0 \leqslant \rho \leqslant 1$ to
ensure that $\mathbf{K}(H)$ is positive definite, hence invertible { and well-conditioned}.

{ To control that the matrix $\mathbf{K}(H)$ is well-conditioned, it is appealing to bound its condition number \citep{won2013condition}. The condition number is the ratio between the largest and the smallest eigenvalue and measures numerical stability: a matrix is well-conditioned if its condition number is moderate, whereas it is ill-conditioned if its condition number is high. To ensure that $\mathbf{K}(H)$ is well-conditioned, it is sufficient to bound the condition number of $\rho \ \mathbf{I} + (1-\rho) c_\alpha \mathbf{S}_{\Wb}(H)$. Since the eigenvalues of $\rho \ \mathbf{I} + (1-\rho) c_\alpha \mathbf{S}_{\Wb}(H)$ equal 
\begin{equation}
\rho + (1-\rho) \lambda,
\label{eq:eigenvalues}
\end{equation}
the  corresponding condition number is 
	\begin{equation}
	CN(\rho) =  \frac{\rho + (1-\rho)\max\{ \lambda \} }{\rho+ (1-\rho)\min\{  \lambda  \}}.
	\label{eq:CN}
	\end{equation}
In practice, we therefore recommend a data-driven approach which sets
$\rho$ at the lowest nonnegative value for which the condition number of $\rho \ \mathbf{I} + (1-\rho) c_\alpha \mathbf{S}_{\Wb}(H)$ is at most $\kappa$. 
This is easy to implement, as we only need to compute the
eigenvalues $\lambda$ of $c_\alpha \mathbf{S}_{\Wb}(H)$ once.
Since regularizing the covariance estimator is our goal and since we mainly focus on very high dimensional data, i.e. situations where $p$ is high compared to the subset size $h$, we recommend prudence and therefore set $\kappa=50$ throughout the paper. This is also the default value in the CovMrcd implementation in the R package {\it rrcov} \citep{rrcov}.}

Note that by this heuristic we only use regularization when
needed. Indeed, if  $\mathbf{S}_{\Wb}(H)$ is well-conditioned,
the heuristic sets $\rho$ equal to zero.
Also note that the eigenvalues in (\ref{eq:eigenvalues})
are at least $\rho$, so the smallest eigenvalue of the MRCD
scatter estimate is bounded away from zero { when $\rho> 0$}.
Therefore the MRCD scatter estimator has a 100\% implosion
 breakdown value { when $\rho> 0$. Note that no affine equivariant scatter estimator can have a breakdown value above 50\% \citep{Lopuhaa:BDP}. The MRCD can achieve this high implosion breakdown value because it is not affine equivariant, unlike the original MCD.}

\section{An algorithm for the MRCD estimator}
\label{Estimation}
\label{estimation}

A naive algorithm for the optimization problem
(\ref{subset2star}) would be to compute $\det(\mathbf{K}(H))$
for every possible $h$-subset $H$.
However, for realistic sample sizes this type of brute
force evaluation is infeasible.

The original MCD estimator (\ref{subset1star}) has the same issue.
The current solution for the MCD consists of either selecting
a large number of randomly chosen initial subsets \citep{RD:99}
or starting from a smaller number of deterministic subsets
\citep{Hubert2012deterministic}.
In either case one  iteratively applies so-called C-steps.
The C-step of MCD improves an $h$-subset $H_1$ by computing
its mean and covariance matrix, and then puts the $h$
observations with smallest Mahalanobis distance in a new
subset $H_2$. The C-step theorem of \citet{RD:99} proves
that the covariance determinant of $H_2$ is lower than or
equal to that of $H_1\;$, so C-steps lower the MCD objective
function.

We will now generalize this theorem to regularized covariance
matrices.

\begin{theorem} \label{cstep}
{Let $\Xb$ be a data set of $n$ points in $p$ dimensions, and take any $n/2 < h < n$ and 
$0 < \rho < 1.$}
Starting from an $h$-subset $H_1,$ one can compute
$\mathbf{m}_1 = \frac{1}{h} \sum_{i \in H_1} \mathbf{x}_i$
and $\mathbf{S}_1 = \frac{1}{h} \sum_{i \in H_1}
(\mathbf{x}_i-\mathbf{m}_1)(\mathbf{x}_i-\mathbf{m}_1)'$.
The matrix
  $$\mathbf{K}_1 = \rho \mathbf{T} +(1-\rho) \mathbf{S}_1$$
is positive definite hence invertible, so we can compute
  $$d_1(i) = (\mathbf{x}_i- \mathbf{m}_1)' \mathbf{K}_1^{-1}
  (\mathbf{x}_i- \mathbf{m}_1)$$
for $i=1,\ldots,n$. Let $H_2$ be an $h$-subset for which
\begin{equation}
  \sum_{i \in H_2} d_1(i) \leq  \sum_{i \in H_1} d_1(i)
	\label{eq:concentration}
\end{equation}
and compute
 $\mathbf{m}_2 = \frac{1}{h} \sum_{i \in H_2} \mathbf{x}_i$,
 $\mathbf{S}_2 = \frac{1}{h} \sum_{i \in H_2}
  (\mathbf{x}_i-\mathbf{m}_2)(\mathbf{x}_i-\mathbf{m}_2)'$ and
 $\mathbf{K}_2 = \rho \mathbf{T} +(1-\rho) \mathbf{S}_2.$
Then
\begin{equation}
  \det(\mathbf{K}_2) \leq  \det(\mathbf{K}_1)
	\label{eq:concresult}
\end{equation}
with equality if and only if $\mathbf{m}_2=\mathbf{m}_1$
and $\mathbf{K}_2=\mathbf{K}_1$.
\end{theorem}

The proof of Theorem \ref{cstep} is given in Appendix A.

 Making use of the generalized C-step we can now construct
the actual algorithm to find the MRCD subset in step 3 of the pseudocode. 

\vskip0.2in
\noindent{\bf ---------------------------------------------------}\\
\vskip-0.4in
\noindent{\bf MRCD algorithm}\\
\vskip-0.4in
\noindent{\bf ---------------------------------------------------}\\
\vskip-0.4in
\begin{enumerate}
\vskip-0.4in
\item [1.] Compute the standardized observations $\ub_i$
		as defined in (\ref{uobs}) using the median and the Qn
		estimator for univariate location and scale.
\item [2.] Perform the singular value decomposition of
    $\mathbf{T}$ into $\mathbf{Q}\mathbf{\Lambda}\mathbf{Q}'$
		where $\mathbf{\Lambda}$ is the diagonal matrix holding the
		eigenvalues of $\mathbf{T}$ and $\mathbf{Q}$ is the
		orthogonal matrix whose columns are the corresponding
		eigenvectors. Compute
		$\wb_i = \mathbf{\Lambda}^{-1/2}\mathbf{Q}'\ub_i\;$.
\item [3.] Find the MRCD subset:
   \begin{enumerate}
   \item [3.1.] Follow Subsection 3.1 in
	   \citet{Hubert2012deterministic} to obtain six robust, well-conditioned initial location
		 estimates $\mb^i$ and scatter estimates
		 $\mathbf{S}^i$ ($i=1,\ldots,6$).
   \item [3.2.] Determine the subsets $H_0^i$ of $\Wb$
	   containing the $h$ observations with lowest
		 Mahalanobis distance in terms of $\mb^i$ and
		 $\mathbf{S}^i$.
   \item [3.3.] For each subset $H_0^i$, determine the smallest
	   value of $0\leq \rho^i<1$ for which
     $\rho^i \ \mathbf{I}+(1-\rho^i)c_\alpha\mathbf{S}_{\Wb}(H_0^i)$
     is well-conditioned. Denote this value as $\rho_0^i\;$.
   \item [3.4.] If $\max_i \rho_0^i \leq 0.1,$ set $\rho =\max_i \rho_0^i,$ else set $\rho =  \max \{0.1; \mbox{median}_i \rho_0^i \} \;$.
   \item [3.5.] For the initial subset $H_0^i$ for which $\rho_0^i \leq \rho$,
				repeat the generalized C-steps from Theorem \ref{cstep} using  $\rho \ \mathbf{I}+(1-\rho)c_\alpha\mathbf{S}_{\Wb}(H_0^i)$
				until convergence. Denote the resulting subsets as $H^i\;$.
   \item [3.6.] Let $H_{MRCD}$ be the subset for which
	      $\rho\ \mathbf{I}+(1-\rho)c_\alpha \mathbf{S}_{\Wb}(H^i)$
				has the lowest determinant among the candidate subsets.
   \end{enumerate}
\item [4.] From $H_{MRCD}$ compute the final MRCD location and
scatter estimates as in (\ref{MRCDestimates}).
\end{enumerate}

In Step 3.1, we first determine the
initial scatter estimates $\Sb^i$ of $\Wb$ in  the same way as
in the DetMCD algorithm of \citet{Hubert2012deterministic}.
This includes the use of steps 4a and 4b of the OGK algorithm of \citet{maronna2002robust} to correct
 for inaccurate eigenvalues and guarantee positive definiteness of the initial estimates. For completeness, the OGK algorithm is provided in Appendix B.
 Given the six initial location and scatter estimates, we then determine in step 3.2 the corresponding six initial subsets of $h$ observations with the lowest Mahalanobis distance. In step 3.3, we compute, for each subset, a regularized covariance, where we use line search and
formula (\ref{eq:eigenvalues}) to calibrate the regularization parameter in such a way that the corresponding condition number
is at most 1000. This leads to potentially six different regularization parameters $\rho_i$.

To ensure comparability of the MRCD covariance estimates on different subsets, we need a unique regularization parameter. In step 3.4, we set by default the final value of the regularization parameter $\rho$ as the largest value of the initial regularization parameters. This is a conservative choice ensuring that the MRCD covariance computed on each subset is well-conditioned. In case of outliers in one of the initial subsets, this may however lead to a too large value
of the regularization parameter. To safeguard the estimation against this outlier inflation of $\rho$, we  change the default choice, when the largest value of all initial $\rho_i$'s exceeds 0.1. We then set the regularization parameter at the median value of the initial regularization parameters, when this median value exceeds 0.1. Otherwise we take 0.1. In the simulation study, we find that in practice $\rho$ tends to be well below 0.1, as long as  the MRCD is implemented with a subset size $h$ that is small enough to resist the outlier contamination. A robust implementation of the MRCD thus ensures that regularization is only used when needed.

In step 3.6, we  recalculate the regularized covariance using $\rho$ instead of $\rho_i$ for each subset with $\rho_i \leq \rho$. We then apply C-steps until the subset no longer changes, which typically requires only
a few steps.
Finally, out of the resulting subsets we select the one
with the lowest objective value, and use it to compute
our final location and scatter estimates according to
(\ref{MRCDestimates}).

\section{Simulation study}
\label{Simulation}

We now investigate the empirical performance of the MRCD.
We compare the MRCD estimator to the OGK estimator of
\citet{maronna2002robust}, which can also robustly estimate
location and scatter in high dimensions but by itself does
not guarantee that the scatter matrix is well-conditioned.
The OGK estimator, as  described in Appendix B,  does not result from optimizing an explicit objective function like the M(R)CD approach. Nevertheless it often works well in practice. { Furthermore, we also compare the MRCD estimator with the RMCD estimator of \citet{croux2012}. We adapted their algorithm to use deterministic instead of random subsets to improve the computation speed. The algorithm that we implemented is described in Appendix C.}

\paragraph{Data generation setup.} In the simulation experiment
we generated $M=500$ contaminated samples of size $n$ from a
$p$-variate normal distribution, with $n \times p$ taken as
either $800\times 100$, $200 \times 100,$ $200\times 200$
and $200\times 400$. Since the MRCD, RMCD and OGK estimators are location and scale equivariant, we follow \citet{agostinelli2015robust}, henceforth ALYZ, by
assuming without loss of generality that the mean $\boldsymbol{\mu}$ is $\mathbf{0}$, and that the diagonal elements of $\mathbf{\Sigma}$ are all equal to unity. As in ALYZ, we account for the lack of affine equivariance of the proposed MRCD estimator by  generating in each
replication the correlation matrix randomly such that the performance of the estimator is not tied to a particular
choice of correlation matrix. We use the  procedure of Section 4 in ALYZ, including the iterative correction to ensure that the condition number of the generated correlation matrix is within a tolerance interval around 100.  To contaminate the data sets, we follow \citet{maronna2002robust} and randomly replace $\lfloor \varepsilon n \rfloor$ observations
by outliers along the eigenvector direction of $\mathbf{\Sigma}$ with smallest eigenvalue, since this is the direction where the contamination is hardest to detect. The distance between the outliers and the mean of the good data is denoted by $k$,
which is set to $50$ for medium-sized outlier contamination and
to $100$ for far outliers.
We let the fraction of contamination $\varepsilon$ be
either 0\% (clean data), 20\% or 40\%.

\paragraph{Evaluation setup.}
On each generated data set we run the MRCD with different
subset sizes $h$, taken as 50\%, 75\%, and 100\% of
the sample size $n$, using the data-driven
choice of $\rho$ with the condition number at most 50. As the target matrix, we take either the identity matrix ($\mathbf{T}=\mathbf{I}_p$) or
the equicorrelation matrix ($\mathbf{T}=\mathbf{R}_c$), with equicorrelation parameter robustly estimated as the average Kendall rank correlation.
{ As non-robust benchmark method we compare with the classical regularized covariance estimator as proposed in \citet{LedoitWolfJMVA}. As robust benchmark method we take the  RMCD using the same subset sizes as used for the MRCD. We also compare with the OGK estimator where the univariate robust scale estimates are obtained using the MAD or the $Q_n$ estimator.}

We measure the inaccuracy of
our scatter estimates $\boldsymbol{S}_m$ compared to the true covariance $\boldsymbol{\Sigma}_m$ by their Kullback-Leiber divergence and mean squared error. { The Kullback-Leiber (KL) divergence measures how much the estimated covariance matrix deviates from the true one by calculating
\[KL(\boldsymbol{S}_m, \boldsymbol{\Sigma}_m)=\text{trace}\left(\boldsymbol{S}_m\boldsymbol{\Sigma}_m^{-1}\right)-\log\left(\det\left(\boldsymbol{S}_m\boldsymbol{\Sigma}_m^{-1}\right)\right)-p\]  } The mean squared error (MSE) is given by
$$MSE=\frac{1}{M}\frac{1}{p^2}\sum_{m=1}^{M}\sum_{k=1}^p
      \sum_{l=1}^{p}\left(\mathbf{S}_m -
			\boldsymbol{\Sigma}_m\right)^2_{k,l}\;\;.$$
Note that the true $\boldsymbol{\Sigma}_m$ differs across
values of $m$ when generating data according to ALYZ.
{ The estimated precision matrices $\boldsymbol{S}_m^{-1}$ are compared by computing their MSE using the true precision matrices $\boldsymbol{\Sigma}_m^{-1}$.

\begin{landscape}

\begin{table}
\label{table:1}
\centering
\caption{Kullback-Leiber and mean squared error of the MRCD, RMCD,  OGK and Ledoit-Wolf scatter matrices for simulation scenarios with 0 and 20\% contamination, together with the MSE of the corresponding precision matrices. }
\vspace{0.5cm}	
\scalebox{0.75}{
\begin{tabular}{l rrrr c rrrr c   rrrr  }
  \hline
   & \multicolumn{4}{c}{\emph{KL} $\hat \Sigma$   }  &  & \multicolumn{4}{c}{\emph{MSE} $\hat \Sigma$   }  & &     \multicolumn{4}{c}{ \emph{MSE} $\hat \Sigma^{-1}$  }   \\
 &  $800\times100$ & $200\times100$ & $200\times200$ & $200\times400$  & &  $800\times100$ & $200\times100$ & $200\times200$ & $200\times400$  & &   $800\times100$ & $200\times100$ & $200\times200$ & $200\times400$  \\ \cline{2-5} \cline{7-10}\cline{12-15}
\multicolumn{10}{l}{\emph{Panel~A: Clean data}  }\\	
MRCD; $ h=\lceil 0.5n \rceil,$ $\mathbf{T}=\mathbf{I}_p$   & 14.8968 & 67.7154 & 255.7381 & 694.7619 &  & 0.0026 & 0.009 & 0.0079 & 0.0083 &  & 0.1143 & 0.3261 & 0.1946 & 0.0728 \\ 
MRCD; $ h=\lceil 0.5n \rceil$, $\mathbf{T}=\mathbf{R}_c$ & 15.7279 & 65.752 & 270.6947 & 669.2409 &  & 0.0026 & 0.009 & 0.0079 & 0.0082 &  & 0.1858 & 0.2969 & 0.2369 & 0.0636 \\ 
MRCD;  $ h=\lceil 0.75n \rceil$,  $\mathbf{T}=\mathbf{I}_p$ & 9.8066 & 42.8 & 172.3765 & 598.0778 &  & 0.0017 & 0.0062 & 0.0056 & 0.0053 &  & 0.0982 & 0.2412 & 0.1635 & 0.0851 \\ 
MRCD; $ h=\lceil 0.75n \rceil$, $\mathbf{T}=\mathbf{R}_c$ & 10.4086 & 41.7183 & 181.4263 & 573.7956 &  & 0.0017 & 0.0062 & 0.0056 & 0.0053 &  & 0.1619 & 0.2202 & 0.2003 & 0.0747 \\ 
MRCD; $ h=n$,  $\mathbf{T}=\mathbf{I}_p$ & 6.8412 & 29.3318 & 122.7992 & 492.7526 &  & 0.0012 & 0.0045 & 0.0042 & 0.0039 &  & 0.0865 & 0.1915 & 0.1282 & 0.0894 \\ 
MRCD;  $h=n$, $\mathbf{T}=\mathbf{R}_c$ & 7.316 & 28.7006 & 129.1631 & 472.403 &  & 0.0012 & 0.0045 & 0.0042 & 0.0039 &  & 0.1441 & 0.1751 & 0.1604 & 0.0782 \\ 
RMCD; $ h=\lceil 0.5n \rceil$ & 112.1142 & 92.7026 & 153.5215 & 343.6736 &  & 0.0045 & 0.0042 & 0.0019 & 9e-04 &  & 0.4521 & 0.3714 & 0.1414 & 0.0636 \\ 
RMCD;   $ h=\lceil 0.75n \rceil$  & 112.2141 & 92.4473 & 153.2346 & 345.7564 &  & 0.005 & 0.0046 & 0.0021 & 9e-04 &  & 0.4537 & 0.3729 & 0.142 & 0.0639 \\ 
RMCD;  $ h=n$  & 112.4177 & 92.3464 & 152.9532 & 346.0013 &  & 0.0057 & 0.0052 & 0.0022 & 0.001 &  & 0.4553 & 0.3743 & 0.1425 & 0.0642 \\ 
  OGK mad  & 8.1344 & 39.7372 & 201.8107 & 1697.4981 &  & 0.0015 & 0.0061 & 0.0058 & 0.0058 &  & 0.0212 & 0.194 & 0.5447 & 4.8724 \\ 
  OGK Qn  & 7.6289 & 36.6829 & 203.0791 & 1803.0049 &  & 0.0014 & 0.0055 & 0.0055 & 0.0056 &  & 0.0176 & 0.1543 & 0.7011 & 6.378 \\ 
  Ledoit-Wolf  & 29.4757 & 57.6959 & 116.5311 & 300.5251 &  & 9e-04 & 0.002 & 0.0013 & 7e-04 &  & 0.3728 & 0.3488 & 0.1356 & 0.0617 \\ 
  \hline
  \multicolumn{10}{l}{\emph{Panel~B: $20\%$ contamination, $k=50$  }  }\\	
MRCD;  $ h=\lceil 0.5n \rceil$,  $\mathbf{T}=\mathbf{I}_p$ & 15.3409 & 68.0281 & 259.0131 & 675.4695 &  & 0.0028 & 0.0104 & 0.0093 & 0.0097 &  & 0.1674 & 0.3474 & 0.2197 & 0.0484 \\ 
MRCD;  $ h=\lceil 0.5n \rceil$, $\mathbf{T}=\mathbf{R}_c$ & 15.1582 & 69.0752 & 241.0183 & 686.5803 &  & 0.0028 & 0.0104 & 0.0093 & 0.0097 &  & 0.1402 & 0.4026 & 0.1468 & 0.0598 \\ 
MRCD; $ h=\lceil 0.75n\rceil$, $\mathbf{T}=\mathbf{I}_p$ & 9.6722 & 41.5235 & 175.003 & 585.7268 &  & 0.0019 & 0.0071 & 0.0065 & 0.0062 &  & 0.1485 & 0.2654 & 0.192 & 0.0606 \\ 
MRCD;  $ h=\lceil 0.75n \rceil$, $\mathbf{T}=\mathbf{R}_c$ & 9.5117 & 42.2133 & 163.5341 & 590.6822 &  & 0.0019 & 0.0071 & 0.0065 & 0.0062 &  & 0.1238 & 0.3158 & 0.1263 & 0.0701 \\ 
MRCD;  $h=n$, $\mathbf{T}=\mathbf{I}_p$ & 223.0005 & 264.1191 & 459.4459 & 599.5314 &  & 1.1061 & 1.0457 & 0.4087 & 0.1643 &  & 0.4037 & 0.4679 & 0.2265 & 0.0408 \\ 
MRCD;  $h=n$, $\mathbf{T}=\mathbf{R}_c$  & 199.8255 & 252.2168 & 389.6361 & 673.7157 &  & 1.1066 & 1.0462 & 0.4098 & 0.1641 &  & 0.3515 & 0.5733 & 0.1274 & 0.0571 \\  
RMCD;  $ h=\lceil 0.5n\rceil$ & 99.1004 & 139.8492 & 183.1091 & 307.896 &  & 0.0108 & 0.0113 & 0.0032 & 0.0013 &  & 0.4209 & 0.6862 & 0.1521 & 0.0598 \\ 
RMCD; $ h=\lceil 0.75n\rceil$  & 99.2097 & 139.7528 & 182.6279 & 306.7708 &  & 0.0126 & 0.0137 & 0.004 & 0.0015 &  & 0.422 & 0.6885 & 0.1529 & 0.0601 \\ 
RMCD;  $h=n$ & 327.937 & 408.0574 & 572.2012 & 1487.1413 &  & 10.6913 & 13.6128 & 8.1852 & 5.5822 &  & 0.4369 & 0.7098 & 0.1613 & 0.0638 \\ 
  OGK mad  & 24.6162 & 72.2043 & 246.5173 & 1043.2791 &  & 0.0077 & 0.0188 & 0.016 & 0.0146 &  & 0.2888 & 0.3813 & 0.1972 & 0.1964 \\ 
  OGK Qn  & 26.004 & 70.703 & 239.0446 & 1001.103 &  & 0.0172 & 0.0339 & 0.0278 & 0.0249 &  & 0.3203 & 0.3986 & 0.1931 & 0.1084 \\
  Ledoit-Wolf & 907.5691 & 637.9635 & 669.0757 & 946.2285 &  & 16.7033 & 14.6159 & 6.9265 & 2.9783 &  & 0.4148 & 0.6968 & 0.1592 & 0.0644 \\ 
  \hline
    \multicolumn{10}{l}{\emph{Panel~C: $20\%$ contamination, $k=100$  }  }\\
MRCD;  $ h=\lceil 0.5n\rceil$,  $\mathbf{T}=\mathbf{I}_p$ & 15.4665 & 68.3158 & 248.8006 & 706.7063 &  & 0.0028 & 0.0104 & 0.0094 & 0.0099 &  & 0.1865 & 0.4573 & 0.1504 & 0.0813 \\ 
MRCD; $ h=\lceil 0.5n\rceil$, $\mathbf{T}=\mathbf{R}_c$ & 14.5347 & 66.7666 & 260.8824 & 704.1722 &  & 0.0028 & 0.0104 & 0.0094 & 0.0099 &  & 0.0881 & 0.3559 & 0.1771 & 0.0839 \\ 
MRCD; $ h=\lceil 0.75n\rceil$, $\mathbf{T}=\mathbf{I}_p$ & 9.8122 & 42.1111 & 168.4095 & 605.9566 &  & 0.0019 & 0.0071 & 0.0066 & 0.0062 &  & 0.1665 & 0.3723 & 0.1293 & 0.09 \\ 
  $ h=\lceil 0.75n \rceil$ T1  & 9.1394 & 41.121 & 176.4705 & 598.3025 &  & 0.0019 & 0.0071 & 0.0066 & 0.0062 &  & 0.0769 & 0.2773 & 0.1534 & 0.0899 \\ 
MRCD;  $h=n$, $\mathbf{T}=\mathbf{I}_p$ & 308.1752 & 296.7125 & 361.9838 & 785.7387 &  & 1.5564 & 1.4755 & 0.5565 & 0.2149 &  & 0.4786 & 0.6612 & 0.1236 & 0.0846 \\ 
MRCD;  $h=n$, $\mathbf{T}=\mathbf{R}_c$  & 217.9974 & 371.0087 & 396.2794 & 845.7801 &  & 1.5517 & 1.4768 & 0.5558 & 0.2149 &  & 0.2825 & 0.5498 & 0.1496 & 0.094 \\ 
RMCD; $ h=\lceil 0.5n\rceil$   & 99.5025 & 64.5539 & 200.7881 & 326.7801 &  & 0.0134 & 0.0097 & 0.0039 & 0.0014 &  & 0.3464 & 0.2456 & 0.1835 & 0.0589 \\ 
RMCD;  $ h=\lceil 0.75n \rceil$ & 99.6584 & 64.5204 & 200.4312 & 325.6408 &  & 0.0155 & 0.0121 & 0.0046 & 0.0015 &  & 0.3476 & 0.2466 & 0.1843 & 0.0591 \\ 
RMCD;  $h=n$ & 627.2463 & 318.8638 & 1057.3902 & 2499.6792 &  & 58.3354 & 69.8824 & 40.9583 & 26.0883 &  & 0.3641 & 0.2588 & 0.1942 & 0.0634 \\ 
  OGK mad  & 25.5102 & 74.8346 & 208.9478 & 1084.1334 &  & 0.009 & 0.0215 & 0.0176 & 0.016 &  & 0.3223 & 0.5318 & 0.1115 & 0.1954 \\ 
  OGK Qn  & 26.6237 & 72.7865 & 202.8054 & 1049.6196 &  & 0.0227 & 0.044 & 0.0341 & 0.0297 &  & 0.3673 & 0.5615 & 0.104 & 0.1225 \\ 
  Ledoit-Wolf  & 1290.1042 & 358.5324 & 838.3355 & 1004.9002 &  & 66.8977 & 58.5843 & 27.7725 & 11.957 &  & 0.3513 & 0.257 & 0.1945 & 0.0648 \\ \hline
 \end{tabular}}
\end{table}
\end{landscape}

\begin{landscape}

\begin{table}
\label{table:2}
\centering
\caption{Kullback-Leiber divergence and Mean squared error of the MRCD, RMCD,  OGK and Ledoit-Wolf  scatter matrices  for simulation scenarios with 40\% contamination, together with the MSE of the corresponding precision matrices. }
\vspace{0.5cm}	
\scalebox{0.75}{
\begin{tabular}{l rrrr c rrrr c   rrrr  }
  \hline
   & \multicolumn{4}{c}{\emph{KL} $\hat \Sigma$   }  &  & \multicolumn{4}{c}{\emph{MSE} $\hat \Sigma$   }  & &     \multicolumn{4}{c}{ \emph{MSE} $\hat \Sigma^{-1}$  }   \\
 &  $800\times100$ & $200\times100$ & $200\times200$ & $200\times400$  & &  $800\times100$ & $200\times100$ & $200\times200$ & $200\times400$  & &   $800\times100$ & $200\times100$ & $200\times200$ & $200\times400$  \\ \cline{2-5} \cline{7-10}\cline{12-15}
   \multicolumn{10}{l}{\emph{Panel~D: $40\%$ contamination, $k=50$  }  }\\	
MRCD;  $ h=\lceil 0.5n \rceil$, $\mathbf{T}=\mathbf{I}_p$ & 14.9182 & 62.9936 & 241.4608 & 663.1413 &  & 0.0032 & 0.012 & 0.0109 & 0.0106 &  & 0.1779 & 0.3117 & 0.1567 & 0.0503 \\ 
MRCD;  $ h=\lceil 0.5n \rceil$, $\mathbf{T}=\mathbf{R}_c$  & 14.86 & 68.928 & 258.7409 & 706.0248 &  & 0.0032 & 0.012 & 0.0109 & 0.0106 &  & 0.1565 & 0.4305 & 0.1901 & 0.0846 \\ 
MRCD;  $ h=\lceil 0.75n\rceil$, $\mathbf{T}=\mathbf{I}_p$ & 324.1335 & 263.4432 & 402.5798 & 654.0979 &  & 1.8617 & 1.7568 & 1.0972 & 0.578 &  & 0.5117 & 0.5016 & 0.1564 & 0.0513 \\ 
 MRCD;  $ h=\lceil 0.75n \rceil$, $\mathbf{T}=\mathbf{R}_c$ & 224.443 & 345.1012 & 481.8516 & 861.7926 &  & 1.8859 & 1.7544 & 1.0965 & 0.5768 &  & 0.3731 & 0.5954 & 0.1863 & 0.0965 \\ 
MRCD;  $h=n$, $\mathbf{T}=\mathbf{I}_p$ & 315.398 & 263.8708 & 394.1002 & 671.6326 &  & 5.5941 & 5.5066 & 2.2665 & 0.9582 &  & 0.5246 & 0.5144 & 0.1593 & 0.052 \\ 
MRCD;  $h=n$, $\mathbf{T}=\mathbf{R}_c$  & 217.9675 & 382.685 & 490.3724 & 885.9611 &  & 5.6189 & 5.5303 & 2.2703 & 0.959 &  & 0.3815 & 0.6067 & 0.1891 & 0.0975 \\ 
RMCD; $ h=\lceil 0.5n\rceil$ & 109.7415 & 76.7225 & 215.2563 & 293.9498 &  & 0.0529 & 0.0459 & 0.0128 & 0.0036 &  & 0.5502 & 0.283 & 0.1883 & 0.05 \\ 
RMCD; $ h=\lceil 0.75n \rceil$  & 116.1911 & 84.9029 & 272.0946 & 622.744 &  & 0.9397 & 1.1853 & 1.0846 & 1.3595 &  & 0.5603 & 0.2924 & 0.196 & 0.0531 \\ 
RMCD;  $h=n$  & 299.6713 & 192.3692 & 561.3463 & 1320.152 &  & 36.4525 & 41.4922 & 24.0353 & 16.4482 &  & 0.5615 & 0.2934 & 0.1965 & 0.0534 \\ 
  OGK mad  & 71.6936 & 107.2329 & 263.0799 & 1020.7997 &  & 0.0408 & 0.077 & 0.0574 & 0.0483 &  & 0.4612 & 0.4556 & 0.1382 & 0.0552 \\ 
  OGK Qn  & 96.0947 & 116.8578 & 258.3033 & 898.1054 &  & 0.1481 & 0.2268 & 0.151 & 0.1239 &  & 0.4891 & 0.4796 & 0.1453 & 0.0453 \\ 
  Ledoit-Wolf & 1332.0961 & 549.9116 & 1163.1474 & 1440.6097 &  & 68.9147 & 63.69 & 31.5632 & 14.6551 &  & 0.5398 & 0.2836 & 0.1924 & 0.0531 \\ 
  \hline
   \multicolumn{10}{l}{\emph{Panel~E: $40\%$ contamination, $k=100$  }  }\\	
MRCD; $h=0.5n$, $\mathbf{T}=\mathbf{I}_p$ & 15.4715 & 67.617 & 254.9857 & 682.7523 &  & 0.0032 & 0.0121 & 0.011 & 0.0108 &  & 0.2385 & 0.4125 & 0.1648 & 0.0608 \\ 
MRCD;  $ h=\lceil 0.5n \rceil$, $\mathbf{T}=\mathbf{R}_c$  & 14.4194 & 66.4281 & 246.5556 & 695.9926 &  & 0.0032 & 0.0121 & 0.011 & 0.0108 &  & 0.1306 & 0.4476 & 0.164 & 0.0683 \\ 
MRCD;  $ h=\lceil 0.75n \rceil$, $\mathbf{T}=\mathbf{I}_p$ & 287.3728 & 341.109 & 478.8436 & 752.5716 &  & 4.9778 & 4.6077 & 2.4917 & 1.1478 &  & 0.635 & 0.5834 & 0.1606 & 0.0651 \\ 
MRCD;  $ h=\lceil 0.75n\rceil$, $\mathbf{T}=\mathbf{R}_c$ & 246.5204 & 419.69 & 472.8565 & 777.6403 &  & 5.1114 & 4.5731 & 2.4749 & 1.1461 &  & 0.3077 & 0.6551 & 0.1744 & 0.0749 \\ 
MRCD;  $h=n$, $\mathbf{T}=\mathbf{I}_p$ & 255.681 & 321.2039 & 451.3078 & 756.7836 &  & 9.9982 & 9.8828 & 3.9458 & 1.6184 &  & 0.6416 & 0.5895 & 0.1622 & 0.0655 \\ 
MRCD;  $h=n$, $\mathbf{T}=\mathbf{R}_c$  & 233.0233 & 408.9508 & 441.2503 & 776.4645 &  & 10.0206 & 9.8887 & 3.9402 & 1.6173 &  & 0.3121 & 0.6617 & 0.1761 & 0.0754 \\ 
RMCD; $ h=\lceil 0.5n \rceil$  & 124.5783 & 85.7827 & 161.004 & 369.5169 &  & 0.0802 & 0.0674 & 0.0175 & 0.0048 &  & 0.5903 & 0.3109 & 0.13 & 0.0801 \\ 
RMCD;  $ h=\lceil 0.75n \rceil$  & 168.0675 & 121.2179 & 303.2048 & 1163.0114 &  & 6.487 & 7.7612 & 6.1159 & 8.7751 &  & 0.6014 & 0.3207 & 0.1359 & 0.0836 \\ 
RMCD;  $h=n$ & 394.9927 & 291.4504 & 736.631 & 2281.4432 &  & 199.7651 & 225.838 & 128.5564 & 80.3043 &  & 0.6024 & 0.3215 & 0.1363 & 0.084 \\ 
  OGK mad  & 62.2365 & 107.4467 & 270.3783 & 1099.2086 &  & 0.062 & 0.1051 & 0.0782 & 0.0665 &  & 0.5784 & 0.5461 & 0.1429 & 0.0642 \\ 
  OGK Qn  & 74.0628 & 113.8013 & 262.9733 & 987.0202 &  & 0.2576 & 0.3707 & 0.255 & 0.2139 &  & 0.6106 & 0.5674 & 0.151 & 0.0579 \\ 
  Ledoit-Wolf  & 1245.4392 & 602.6419 & 1133.525 & 1730.1454 &  & 275.9802 & 255.161 & 126.4966 & 58.7499 &  & 0.5886 & 0.3165 & 0.135 & 0.0842 \\ 
   \hline
\end{tabular}}
\end{table}
\end{landscape}

\clearpage

\begin{table}
\label{table:3}
\centering
\caption{Average value of $\rho$,
across 500 replications of the ALYZ data generating
process.}
\vspace{0.5cm}	
\scalebox{0.85}{
\begin{tabular}{rrrrr c rrrrr}
  \hline
 &  \multicolumn{4}{c}{$\mathbf{T}=\mathbf{I}_p$}&& \multicolumn{4}{c}{$\mathbf{T}=\mathbf{R}_c p$}\\
 & 800x100 & 200x100 & 200x200 & 200x400 & &800x100 & 200x100 & 200x200 & 200x400 \\
 \cline{2-5} 	\cline{7-10} 
\multicolumn{10}{l}{\emph{Panel~A: Clean data}  }\\	
 $ h=\lceil  0.5n \rceil$  & 0.05 & 0.09 & 0.12 & 0.18 &&  0.05 & 0.09 & 0.12 & 0.18 \\ 
  $ h=\lceil  0.75n\rceil$  & 0.04 & 0.07 & 0.10 & 0.14 && 0.04 & 0.07 & 0.10 & 0.14 \\  
  $h=n$  & 0.03 & 0.06 & 0.09 & 0.12 & & 0.03 & 0.06 & 0.09 & 0.12 \\ \hline
    \multicolumn{10}{l}{\emph{Panel~B: $20\%$ contamination, $k=50$  }  }\\	
   $ h=\lceil h=0.5n \rceil$ & 0.02 & 0.05 & 0.06 & 0.10 & & 0.02 & 0.05 & 0.06 & 0.10 \\ 
  $ h=\lceil 0.75n \rceil$ & 0.02 & 0.04 & 0.05 & 0.07 & & 0.02 & 0.04 & 0.05 & 0.07 \\ 
  $h=n$& 0.76 & 0.78 & 0.82 & 0.85 & & 0.76 & 0.78 & 0.82 & 0.85 \\ \hline
      \multicolumn{10}{l}{\emph{Panel~C: $20\%$ contamination, $k=100$  }  }\\
  $ h=\lceil 0.5n \rceil$ & 0.02 & 0.04 & 0.06 & 0.09 & & 0.02 & 0.04 & 0.06 & 0.09 \\ 
  $ h=\lceil 0.75n\rceil$  & 0.02 & 0.04 & 0.05 & 0.07 & & 0.02 & 0.04 & 0.05 & 0.07 \\ 
  $h=n$ & 0.86 & 0.87 & 0.89 & 0.92 & & 0.86 & 0.87 & 0.89 & 0.92 \\ \hline
      \multicolumn{10}{l}{\emph{Panel~D: $40\%$ contamination, $k=50$  }  }\\	
  $ h=\lceil 0.5n \rceil$ & 0.01 & 0.02 & 0.03 & 0.04 & & 0.01 & 0.02 & 0.03 & 0.04 \\ 
   $ h=\lceil 0.75n \rceil$ & 0.29 & 0.35 & 0.50 & 0.63 & & 0.29 & 0.34 & 0.50 & 0.63 \\ 
  $h=n$ & 0.73 & 0.74 & 0.78 & 0.81 & & 0.73 & 0.74 & 0.78 & 0.81 \\ \hline
   \multicolumn{10}{l}{\emph{Panel~E: $40\%$ contamination, $k=100$  }  }\\	
   $ h=\lceil 0.5n\rceil$  & 0.01 & 0.02 & 0.02 & 0.04 & & 0.01 & 0.02 & 0.02 & 0.04 \\ 
   $ h=\lceil 0.75n\rceil$  & 0.43 & 0.47 & 0.62 & 0.74 & & 0.40 & 0.47 & 0.62 & 0.74 \\ 
  $h=n$ & 0.82 & 0.82 & 0.85 & 0.88 & & 0.82 & 0.82 & 0.85 & 0.88 \\
   \hline
\end{tabular}}
\end{table}

\paragraph{Discussion of results.}
The results are reported in Tables 1 and 2. Table 1 presents the simulation scenarios in the absence of outlier contamination and with $20\%$ contamination, while  Table 2 shows the results when there are $40\%$ outliers present in the data. The left panel shows the MSE of the scatter matrices, the middle panel lists the KL divergence of the scatter matrices and the right panel reports the MSE of the precision matrices. 

In terms of the MSE and the KL divergence of the covariance estimates we find that, in the case of no outlier contamination, the MRCD covariance estimate with $h=n$ has the lowest MSE when $n>p$. The RMCD estimators perform worse in this situation. If $p$ becomes bigger than $n$, the classical regularized covariance estimator performs the best, closely followed by RMCD and MRCD. Note that for these situations, the OGK estimator has clearly the weakest performance. The performance of the MRCD estimator with $h=\lceil 0.5 n \rceil$ is clearly less than the MRCD estimator with $h=n$. This lower efficiency is compensated by the high breakdown robustness. In fact, for both 20\% and 40\% outlier contamination, the MSE and KL divergence of the MRCD with $h=\lceil 0.5 n \rceil$ is very similar to the one in the absence of outliers, and it is always substantially lower than the MSE of the OGK covariance estimator.

When outliers are added to the data, the Ledoit-Wolf  covariance matrix and the MRCD and RMCD estimators with $h=n$ immediately break down. As expected, the MRCD and RMCD estimators with $h=\lceil 0.75n\rceil$ perform best when there is $20\%$ contamination and $h=\lceil 0.5n\rceil$ is the only reliable choice when there are $40\%$ of outliers in the data. Note that our proposed estimators outperform the OGK estimator in every situation.

Similar conclusions can be drawn for the performance of the estimated precision matrices. The MRCD and RMCD precision estimates  both remain accurate in the presence of outliers as long as the subsample size $h$ does not exceed the number of clean observations.

The simulation study also sheds light on how the structure of the data and the presence of outlier contamination affect the calibration of the regularization parameter $\rho$. Table 3 lists the average value of the data-driven $\rho$ for the MRCD covariance estimator. Recall that the MRCD  uses the smallest value of $0\leqslant\rho<1$ for which the scatter matrix is well-conditioned, so when the MCD is well-conditioned the MRCD  obtains $\rho=0$
and thus coincides with the MCD in that case.  We indeed find that $\rho$ is close to $0$ in the scenarios where $h>p$ and  $h < n (1- \epsilon)$, and that $\rho$ remains close to zero when the subset size $h$ is small enough to resist the outlier contamination. It follows that the choice between the identity matrix or the robustly calibrated equicorrelation matrix as target matrix has only a negligible impact on the MSE, provided the MRCD is implemented with a subset size $h$ that is small enough to resist the outlier contamination. When the number of outliers exceeds the subset size, we see that outliers induce higher $\rho$ values.

In conclusion, the simulation study confirms that the MRCD
is a good method for estimating location and scatter in high
dimensions.
It only regularizes when needed.
When $h$ is less than $p$ and the number of clean observations,
the resulting $\rho$ is typically less than {10\%}, implying
that the MRCD strikes a balance between being similar
to the MCD for tall data and achieving a well-conditioned
estimate in the case of fat data.}

\section{Real data examples}
\label{Applications}

We illustrate the MRCD on two datasets with low $n/p$,
so using the original MCD is not indicated. The MRCD is implemented using the identity matrix as target matrix.

\subsection{Octane data}

The octane data set described in
\citet{esbensen1996multivariate} consists of near-infrared
absorbance spectra with $p=226$ wavelengths collected on
$n=39$ gasoline samples.
It is known that the samples 25, 26, 36, 37, 38 and 39 are
outliers which contain added ethanol \citep{Hubert:ROBPCA}.
Of course, in most applications the number of outliers is not
known in advance hence it is not obvious to set the subset
size $h$.
The choice of $h$ matters because increasing $h$ improves
the efficiency at uncontaminated data but hurts the
robustness to outliers.
Our recommended default choice is $h=\lceil 0.75 n \rceil$, safeguarding the MRCD covariance estimate against up to
$25\%$ of outliers.

\begin{figure}[ht!]
\centering
\begin{minipage}[l]{0.45\textwidth}
	\centering
	\includegraphics[width=0.9\textwidth]{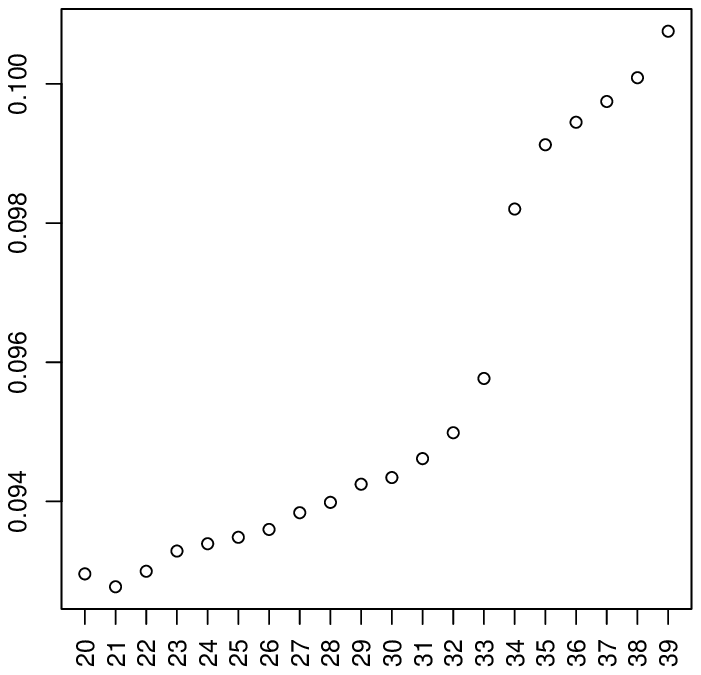}
	\caption{Octane data: MRCD objective value (\ref{subset2star1})
	           for different values of $h$.}
	\label{fig:octaneh}
\end{minipage}
\hfill
\begin{minipage}[l]{0.45\textwidth}
	\centering
		\includegraphics[width=0.9\textwidth,angle=0]{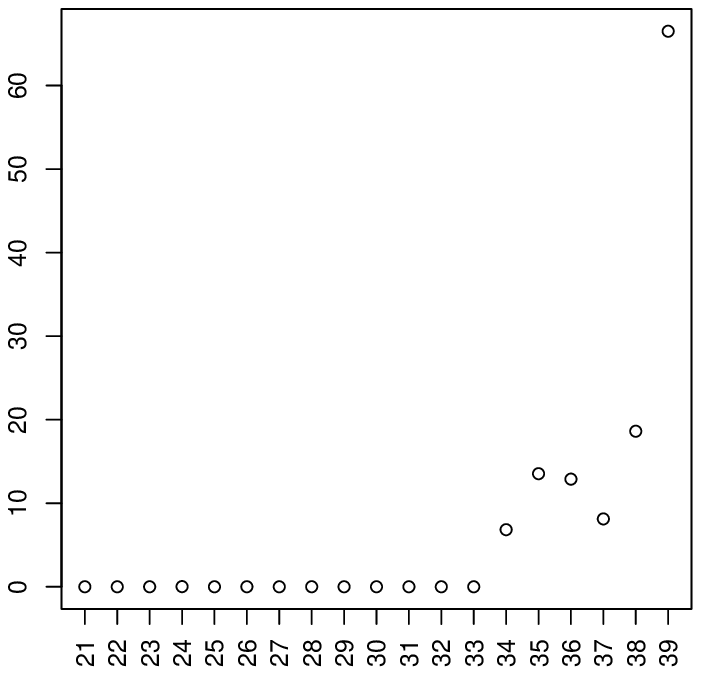}
	\caption{Octane data: Frobenius distance between MRCD scatter matrices on standardized data
	        for $h-1$ and $h$.}
	\label{fig:octanedisth}
\end{minipage}
\hfill
\begin{minipage}[l]{0.8\textwidth}
	\includegraphics[width=1\textwidth]{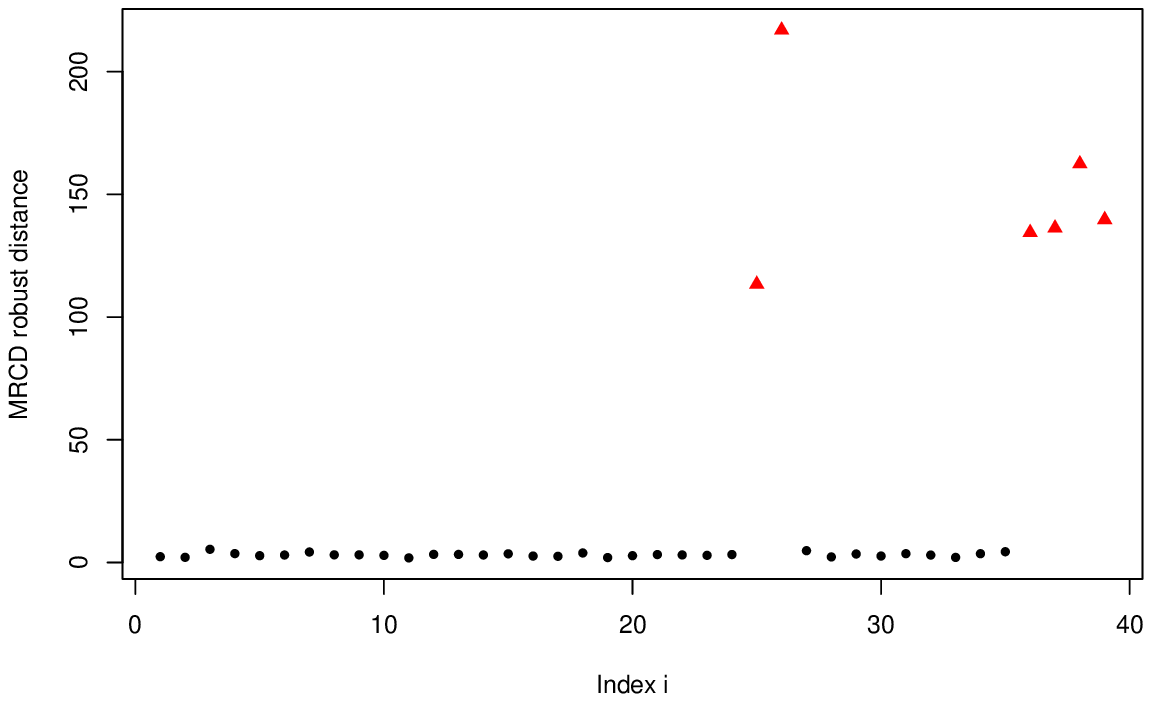}
	\caption{Robust distances of the octane data, based on the
	         MRCD with $h=33$.}
	\label{octanedist}
\end{minipage}
\end{figure}

Alternatively, one could employ a data-driven approach to
select $h$. {This idea is similar to the forward search of \citet{ForwardSearch}.}
It consists of computing the MRCD for a range of $h$ values,
and looking for an important change in the objective function
or the estimates at some value of $h$.
This is not too hard, since we only need to
obtain the initial estimates $\Sb^i$ once.
Figure \ref{fig:octaneh} plots the MRCD objective function
(\ref{subset2star1}) for each value of $h$, while Figure \ref{fig:octanedisth} shows the Frobenius distance
between the  MRCD scatter matrices of the standardized data (\emph{i.e.}, $\rho \ \mathbf{I} + (1-\rho)\mathbf{S}_{\Wb}(H_{MRCD}) )$, as defined in (\ref{MRCDestimates})) obtained for $h-1$ and $h$.
Both figures clearly indicate that there is an important
change at $h=34$, so we choose $h=33\;$.
The total computation time to produce these plots was only 12 seconds on an Intel(R) Core(TM) i7-5600U CPU with 2.60 GHz.

We then calculate the MRCD estimator with $h=33$, yielding
$\rho=0.1149$.
Figure \ref{octanedist} shows the corresponding robust
distances
\begin{equation}
RD(\xb_i)=\sqrt{(\xb_i-\mathbf{m}_{MRCD})'\mathbf{K}_{MRCD}^{-1}
          (\xb_i-\mathbf{m}_{MRCD})}
\end{equation}
where $\mathbf{m}_{MRCD}$ and $\mathbf{K}_{MRCD}$ are the MRCD
location and scatter estimates of (\ref{MRCDestimates}).
The flagged outliers (red triangles) stand out, showing the
MRCD has correctly identified the 6 samples with added ethanol.

\subsection{Murder rate data}

\citet{khan2007robust} regress the murder rate per 100,000
residents in the $n=50$ states of the US in 1980 on 25
demographic predictors, and mention that graphical tools
reveal one clear outlier.

For lower-dimensional data, \citet{Rousseeuw:MCDreg}
applied the MCD estimator to the response(s) and predictors
together to robustly estimate a multivariate regression.
Here we investigate whether for high-dimensional data
the same type of analysis can be carried out based on the MRCD.
In the murder rate data this yields a total of 26 variables.

\begin{figure}[ht!]
\centering
\begin{minipage}[l]{0.45\textwidth}
	\centering
	\includegraphics[width=1\textwidth]{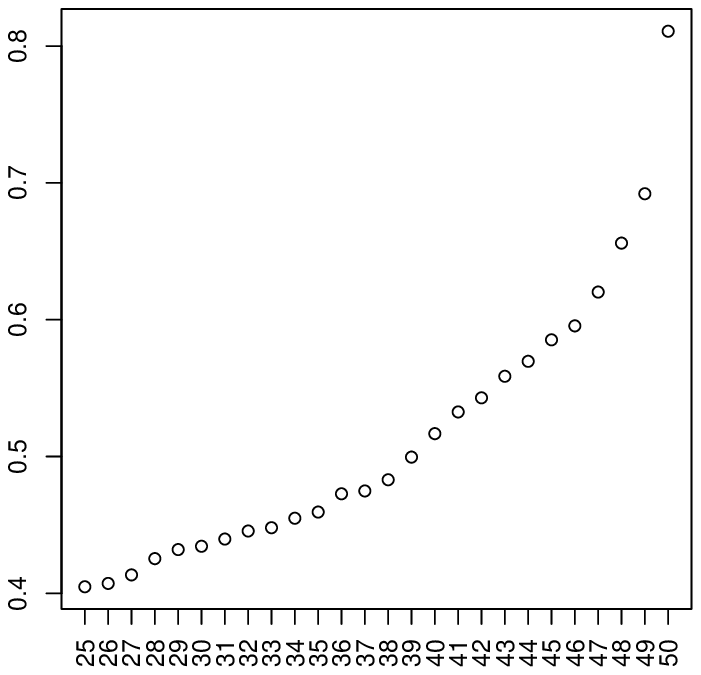}
	\caption{Murder rate data: MRCD objective value (\ref{subset2star1})
					 for different values of $h$.}
	\label{fig:regh}
\end{minipage}
\hfill
\begin{minipage}[l]{0.45\textwidth}
	\centering
		\includegraphics[width=1\textwidth]{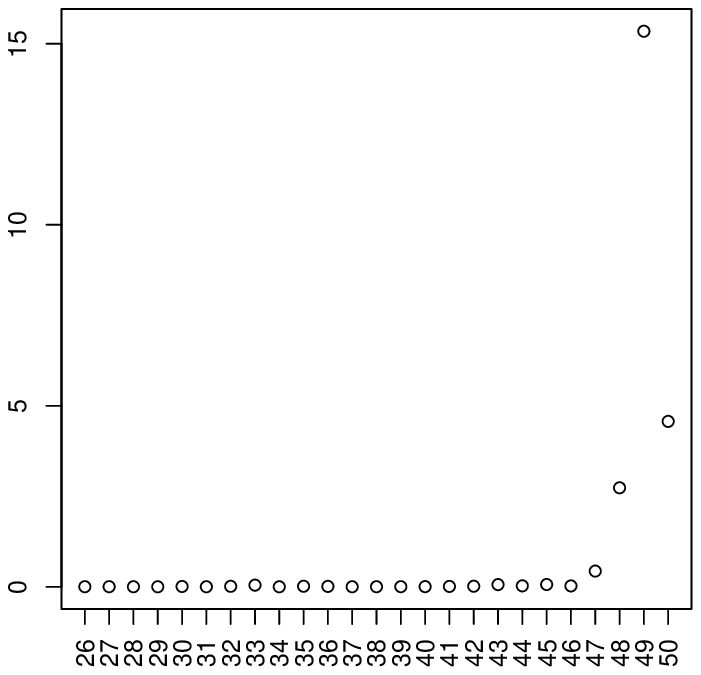}
	\caption{Murder rate data:  Frobenius distance between MRCD scatter matrices on standardized data
					  for $h-1$ and $h$.}
	\label{fig:regdisth}
\end{minipage}
\end{figure}

As for the octane data, we compute the MRCD estimates for
the candidate range of $h$.
In Figure \ref{fig:regh} we see a big jump in the objective
function when going from $h=49$ to $h=50$.
But in the plot of the Frobenius distance between successive
MRCD scatter matrices (Figure \ref{fig:regdisth}) we see
evidence of  four outliers, which lead to a substantial change
in the MRCD when included in the subset.

As a conservative choice we set $h=44$,
which allows for up to 6 outliers.
We then partition the MRCD scatter matrix on all 26 variables
as follows:
 $$ \mathbf{K}_{MRCD} = \left( \begin{array}{cc}
 \mathbf{K}_{xx} &  \mathbf{K}_{xy} \\ \mathbf{K}_{xy} &
 \mathbf{K}_{yy} \end{array} \right) ,$$
where $x$ stands for the vector of predictors and $y$ is the
response variable.
The resulting estimate of the slope vector is then
 $$\hat \beta_{MRCD} = \mathbf{K}_{xx}^{-1} \mathbf{K}_{xy}\;.$$

\begin{figure}[ht!]
\centering
\includegraphics[width=0.85\textwidth]{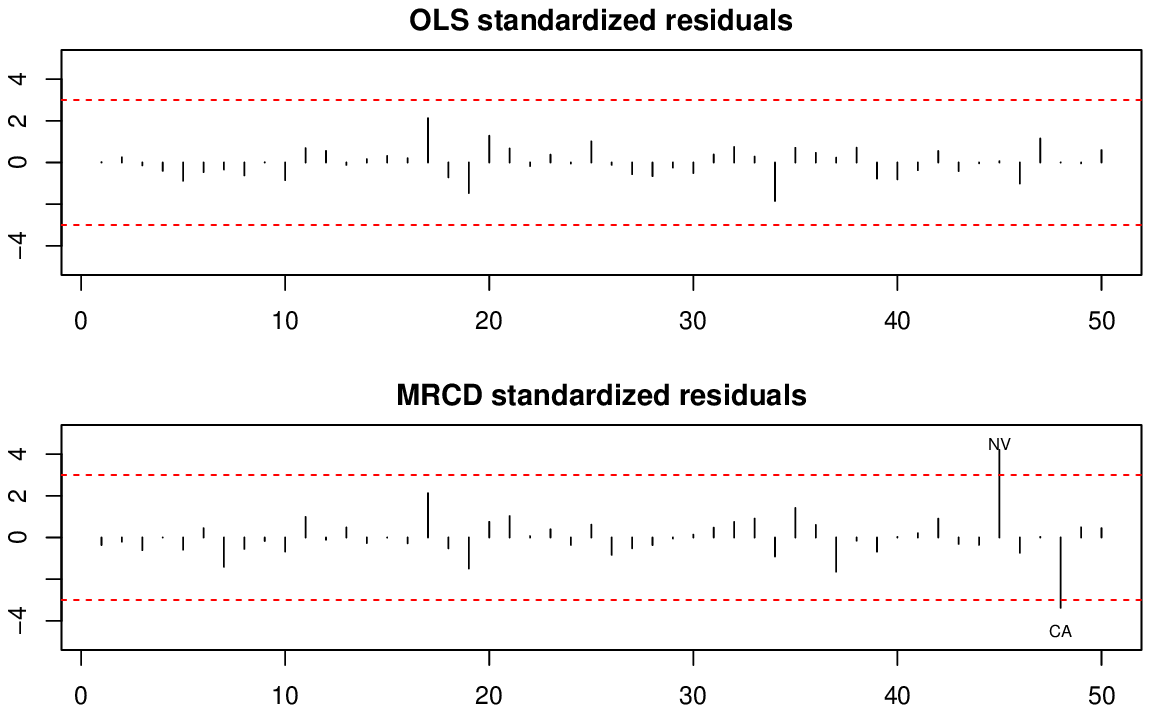}
\caption{Standardized residuals of regressing
         the murder rate on demographic variables.}
\label{fig:beta}
\end{figure}

The resulting standardized residuals are shown in
Figure \ref{fig:beta}. { The standardized residuals obtained with OLS indicate that there are no outliers in the data since all residuals are clearly between the cut-off lines. In contrast, the MRCD regression flags \textit{Nevada} as an upwards outlier and \textit{California} as a downwards outlier. It is therefore  recommended to study these states in more detail. Note that both states have very small residuals when using OLS. This is a clear example of the well known masking effect: classical methods can be affected by outliers so strongly that the resulting fitted model does not allow to detect the deviating observations. 

Finally, we note that MRCD regression can be plugged into
existing robust algorithms for variable selection, which
avoids the limitation mentioned in \citet{khan2007robust}
that ``a robust fit of the \textit{full} model may not be
feasible due to the numerical complexity of robust estimation
when [the dimension] $d$ is large (e.g., $d$ $\geq$ 200) or
simply because $d$ exceeds the number of cases, $n$.''
The MRCD could be used in such situations because its
computation remains feasible in higher dimensions.
}
\section{Concluding remarks}
\label{Conclusions}

In this paper we generalized the Minimum Covariance Determinant (MCD) estimation approach of \citet{rousseeuw1985multivariate} to
higher dimensions, by regularizing the sample covariance
matrices of subsets before minimizing their determinant.
The resulting Minimum Regularized Covariance Determinant
(MRCD) estimator is well-conditioned by construction, even
when $p>n$, and preserves the good robustness of the MCD.
We constructed a fast algorithm for the MRCD by
generalizing the C-step used by the MCD, and proving that
this generalized C-step is guaranteed to reduce the
covariance determinant.
We verified the performance of the MRCD estimator in an
extensive
simulation study including both clean and contaminated data.
The simulation study also confirmed that the MRCD can be
interpreted as a generalization of the MCD. When
$n$ is sufficiently large compared to $p$ and the MCD is
well-conditioned, the regularization parameter in MRCD
becomes zero and the MRCD estimate coincides with the MCD.
Finally, we illustrated the use of the MRCD for outlier
detection and robust regression on two fat data applications
from chemistry and criminology, for which $p>n/2$.

We believe that the MRCD is a valuable addition to the tool set
for robust multivariate analysis, especially in high dimensions. 
{Thanks to the function CovMrcd in the R package {\it rrcov} of \citet{rrcov}, practitioners and academics can easily implement our methodology in practice.} We look forward to further research on its use in principal
component analysis where the original MCD has proved
useful~\citep{Croux:PCAMCD,Hubert:ROBPCA}, and analogously
in factor analysis~\citep{Pison:RobFA},
classification~\citep{Hubert:Discrim},
clustering~\citep{Hardin:ClusteringMCD},
multivariate regression~\citep{Rousseeuw:MCDreg},
penalized maximum likelihood estimation \citep{croux2012}
and other multivariate techniques.
A further research topic is to study the finite sample
distribution of the robust distances computed from the MRCD.
Our experiments have shown that the usual chi-square and F-distribution results for the MCD distances
\citep{hardin2005distribution} are no longer good
approximations when $p$ is large relatively to $n$.
A better approximation would be useful for improving the
accuracy of the MRCD by reweighting.


\vskip0.4in
\large
\noindent {\bf Appendix A: Proof of Theorem \ref{cstep}}
\normalsize
\vskip0.1in
\noindent Generate a $p$-variate sample $\mathbf{Z}$ with
$p+1$ points for which
$\mathbf{\Lambda} = \frac{1}{p+1}\sum_{j=1}^{p+1}
 (\zb_i-\overline{z})(\zb_i-\overline{z})'$ is nonsingular
and\linebreak
$\overline{z}=\frac{1}{p+1}\sum_{j=1}^{p+1}\zb_i$. Then $\tilde{\zb}_i=\mathbf{\Lambda}^{-1/2}(\zb_i-\overline{z})$ has mean zero and covariance matrix $\mathbf{I}_p$.
Now compute $\yb_i=\mathbf{T}^{1/2}\tilde{\zb}_i\;$, hence
$\mathbf{Y}$ has mean zero and covariance matrix $\mathbf{T}$.

Next, create the artificial dataset
\[ \tilde{\mathbf{X}}^{1} = \left( w_1(\xb^{1}_{1}-\mathbf{m}_1),\ldots,w_h(\xb^{1}_{h}-\mathbf{m}_1),w_{h+1}\yb_1,\ldots,w_{k}\yb_{p+1}\right) \]
with $k=h+p+1$ points, where $\xb^{1}_{1},\ldots,\xb^{1}_{h}$
are the members of $H_1$.
The factors $w_i$ are given by
$$w_i =  \left\{ \begin{array}{cl} \sqrt{k(1-\rho)/h} & \;\;\; \mbox{for} \ i=1,\ldots,h \\ \sqrt{k\rho/(p+1)} & \;\;\; \mbox{for} \  i=h+1,\ldots,k\;\;. \end{array}\right.$$
The mean and covariance matrix of $ \mathbf{\tilde X}^1$ are then
\begin{align*}
\frac{1}{k} \sum_{i=1}^k  \boldsymbol{\tilde x}^1_i &= \sqrt{\frac{1-\rho}{kh}} \sum_{i=1}^h ( \boldsymbol{x}^1_i - \mathbf{m}_1) +\sqrt{\frac{\rho}{k(p+1)}} \sum_{j=1}^{p+1}  \yb_j = 0
\end{align*}
and
\begin{align*}\frac{1}{k} \sum_{i=1}^k  \boldsymbol{\tilde x}^{1}_{i} (\boldsymbol{\tilde x}^{1}_{i})' &=
\frac{1-\rho}{h} \sum_{i=1}^h ( \boldsymbol{x}^{1}_{i} - \mathbf{m}_1)( \boldsymbol{x}^{1}_{i} - \mathbf{m}_1)'
+\frac{\rho}{p+1}  \sum_{j=1}^{p+1}  \yb_j \yb'_j \\
&= (1-\rho) \mathbf{S}_1 + \rho \mathbf{T} = \mathbf{K}_1\;\;.
\end{align*}
The regularized covariance matrix $\mathbf{K}_1$ is thus the
actual covariance matrix of the combined data set
$\mathbf{\tilde{X}}^1\;$.
Analogously we construct
$$ \tilde{\mathbf{X}}^2 = \left( w_1(\xb^{2}_{1}-\mathbf{m}_2),\ldots,w_h(\xb^{2}_{h}-\mathbf{m}_2),w_{h+1}\yb_1,\ldots,w_{k}\yb_{p+1}\right) $$
where $\xb^{2}_{1},\ldots,\xb^{2}_{h}$ are the members of $H_2\;$.
$\tilde{\mathbf{X}}_2$ has zero mean and covariance matrix
$\mathbf{K}_2=(1-\rho) \mathbf{S}_2 + \rho \mathbf{T}\;$.

Denote $d_{\mathbf{K}_1}(\boldsymbol{\tilde x}) =
 \boldsymbol{\tilde x'}( \mathbf{K}_1)^{-1}\boldsymbol{\tilde x}$.
We can then prove that:
\begin{align}
\frac{1}{k}\sum_{i=1}^h d_{\mathbf{K}_1}(\boldsymbol{\tilde x}^2_{i} ) &=\frac{1-\rho}{h}\sum_{i=1}^h d_{\mathbf{K}_1}(\xb^2_{i}-\mathbf{m}_2) \\
&\leq \frac{1-\rho}{h}\sum_{i=1}^h d_{\mathbf{K}_1}(\xb^2_{i}-\mathbf{m}_1) \label{eq:a2}\\	
&\leq \frac{1-\rho}{h}\sum_{i=1}^h d_{\mathbf{K}_1}(\xb^1_{i}-\mathbf{m}_1) \label{eq:a3}\\
&= \frac{1}{k}\sum_{i=1}^h d_{\mathbf{K}_1}(\mathbf{\tilde x}^1_{i})
\end{align}
in which the second inequality (\ref{eq:a3}) is the
condition (\ref{eq:concentration}).

The first inequality (\ref{eq:a2}) can be shown as follows.
Put $\zb_i = (\mathbf{K}_1)^{-1/2}\xb^2_{i}$ and $\tilde{\zb} = (\mathbf{K}_1)^{-1/2}\mathbf{m}_1$ and note that $\overline{\zb}=(\mathbf{K}_1)^{-1/2}\mathbf{m}_2$ is the average of the $\zb_i$. Then (\ref{eq:a2}) becomes
$$\sum_{i=1}^h \| \zb_i -\overline{\zb}   \|^2 \leq  \sum_{i=1}^h \| \zb_i - \tilde{\zb}   \|^2,$$
which follows from the fact that $\tilde{\zb} $ is the unique minimizer of the least squares objective $\sum_{i=1}^k \| \zb_i - c  \|^2$, so (\ref{eq:a2}) becomes an equality if and only if
$\tilde{\zb} =\overline{\zb} $ which is equivalent to
$\mathbf{m}_2=\mathbf{m}_1$.\\
It follows that
\begin{align*}
\sum_{i=1}^k d_{\mathbf{K}_1}(\boldsymbol{\tilde x}^2_{i} ) &=\sum_{i=1}^h d_{\mathbf{K}_1}(\boldsymbol{\tilde x}^2_{i} ) + \frac{\rho}{p+1}\sum_{j=1}^{p+1} d_{\mathbf{K}_1}(\yb_{j} ) \\
&\leq \sum_{i=1}^h d_{\mathbf{K}_1}(\boldsymbol{\tilde x}^1_{i} ) + \frac{\rho}{p+1}\sum_{j=1}^{p+1} d_{\mathbf{K}_1}(\yb_{j} ) \\
&= \sum_{i=1}^k d_{\mathbf{K}_1}(\boldsymbol{\tilde x}^1_{i})\;\;.
\end{align*}
Now put
\[ b = \frac{ \sum_{i=1}^k d_{\mathbf{K}_1}(\boldsymbol{\tilde x}^2_{i} ) }{ \sum_{i=1}^k d_{\mathbf{K}_1}(\boldsymbol{\tilde x}^1_{i} )}  \leq 1\;\;.\]
If we now compute distances relative to $b \mathbf{K}_1\;$,
we find
\begin{align*}
\lefteqn{\frac{1}{k} \sum_{i=1}^k d_{b\mathbf{K}_1}(\boldsymbol{\tilde x}^2_{i} )=\frac{1}{b} \frac{1}{k} \sum_{i=1}^k d_{\mathbf{K}_1}(\boldsymbol{\tilde x}^2_{i} ) =  \frac{1}{k} \sum_{i=1}^k d_{\mathbf{K}_1}(\boldsymbol{\tilde x}^1_{i} ) = \frac{1}{k} \sum_{i=1}^k (\boldsymbol{\tilde x}^{1}_{i} )'(\mathbf{K}_1)^{-1} \boldsymbol{\tilde x}^{1}_{i}}\\
&=\frac{1}{k} \sum_{i=1}^k (\mathbf{K}_1^{-1/2}\boldsymbol{\tilde x}^{1}_{i} )'(\mathbf{K}_1^{-1/2}\boldsymbol{\tilde x}^{1}_{i})= \mbox{Trace}\left(  \frac{1}{k} \sum_{i=1}^k (\mathbf{K}_1^{-1/2}\boldsymbol{\tilde x}^{1}_{i} )'(\mathbf{K}_1^{-1/2}\boldsymbol{\tilde x}^{1}_{i})  \right) \\
&= \mbox{Trace}\left( (\mathbf{K}_1)^{-1/2} \left( \frac{1}{k} \sum_{i=1}^k (\boldsymbol{\tilde x}^{1}_{i} )(\boldsymbol{\tilde x}^{1}_{i})' \right)(\mathbf{K}_1)^{-1/2}  \right) = \mbox{Trace}(\mathbf{I}_p) = p\;\;.
\end{align*}
From the theorem in \citet{grubel1988minimal}, it follows that
$\mathbf{K}_2$ is the unique minimizer of
$\mbox{det}(\mathbf{S})$ among all $\mathbf{S}$ for which
$\frac{1}{k} \sum_{i=1}^k
 d_{\mathbf{S}}(\boldsymbol{\tilde x}^{2}_{i} )=p$
(note that the mean of $\boldsymbol{\tilde x}^{2}_{i}$ is zero).
Therefore
$$\det(\mathbf{K}_2) \leq  \det(b\mathbf{K}_1) \leq  \det(\mathbf{K}_1)\;\;.$$
We can only have $\mbox{det}(\mathbf{K}_2) = \det(\mathbf{K}_1)$
if both of these inequalities are equalities.
For the first, by\;\; uniqueness we can only have equality
if $\mathbf{K}_2=b\mathbf{K}_1$.
For the second inequality, equality holds if and only if $b=1$. Combining both yields $\mathbf{K}_2=\mathbf{K}_1$.
Moreover, $b=1$ implies that (\ref{eq:a2}) becomes an equality,
 hence $\mathbf{m}_2=\mathbf{m}_1$.
This concludes the proof of Theorem \ref{cstep}.\\

\vskip0.2in
\large
\noindent {\bf Appendix B: The OGK estimator}\\
\normalsize

\noindent \citet{maronna2002robust} presented a general method to obtain positive definite
and approximately affine equivariant robust scatter matrices starting from a robust bivariate scatter measure.
This method was applied to the bivariate covariance estimate
of~\citet{gnanadesikan1972robust}.
The resulting multivariate location and scatter estimates are called
orthogonalized Gnanadesikan-Kettenring (OGK) estimates and are calculated
as follows:
\begin{enumerate}
\item Let $m(.)$ and $s(.)$ be robust univariate estimators of location
      and scale.
\item Construct $\yb_i=\boldsymbol{D}^{-1}\xb_i$ for $i=1,\ldots,n$ with
      $\boldsymbol{D}=\text{diag}(s(X_1),\ldots,s(X_p))\;$.
\item Compute the `pairwise correlation matrix' $\Ub$ of
      the variables of $\Yb=(Y_1,\ldots,Y_p)\;$, given by
      $u_{jk} = 1/4 (s(Y_j+Y_k)^2-s(Y_j-Y_k)^2)\;$.
	    This $\Ub$ is symmetric but not necessarily positive
			definite.
\item Compute the matrix $\Eb$ of eigenvectors of $\Ub$ and
\begin{enumerate}
\item project the data on these eigenvectors,
      i.e. $\Vb=\Yb\Eb\;$;
\item compute `robust variances' of $\Vb=(V_1,\ldots,V_p)\;$, i.e.
      $\boldsymbol{\Lambda} =
			\text{diag}(s^2(V_1),\ldots,s^2(V_p))\;$;
\item set the $p \times 1$ vector $\hm(\Yb) = \Eb \mb$ where
      $\mb=(m(V_1),\ldots,m(V_p))^T\;$,
      and compute the positive definite matrix
      $\hS(\Yb) = \Eb \boldsymbol{\Lambda} \Eb^T\;$.
\end{enumerate}
\item Transform back to $\Xb$, i.e. $\hmrewo = \Db \hm(\Yb)$
      and $\hSrewo = \Db \hS(\Yb) \Db^T\;$.
\end{enumerate}
Step 2 makes the estimate location invariant and scale
equivariant, whereas the next steps replace the eigenvalues
of $\Ub$ (some of which may be negative) by positive numbers. In the simulation study and empirical analysis, we set $m(.)$ to the median and $s(.)$ to {either } the median absolute deviation {or the Qn scale estimator. We use the implementation in the R package {\it rrcov} of \citet{rrcov}}. 

\vskip0.2in
\large
\noindent {\bf Appendix C: The RMCD estimator}\\
\normalsize

The RMCD as initially proposed by \citet{croux2012} uses random subsets. Below we give its adaptation using
deterministic subsets. {We thank Christophe Croux and Gentiane Haesbrouck for their helpful guidelines in specifying the proposed detRMCD algorithm in which we follow closely the MRCD algorithm presented in Section \ref{estimation}.
} It uses the GLASSO algorithm of \citet{glasso2008}, as implemented in the package {\it huge}  of \citet{zhao2012huge}.

\vskip0.2in
\noindent{\bf ---------------------------------------------------}\\
\vskip-0.4in
\noindent{\bf (det)RMCD algorithm}\\
\vskip-0.4in
\noindent{\bf ---------------------------------------------------}\\
\vskip-0.4in
\begin{enumerate}
\vskip-0.4in
\item [1.] Compute the standardized observations $\ub_i$
		as defined in (\ref{uobs}) using the median and the Qn
		estimator for univariate location and scale.

\item [2.] Find the RMCD subset:
   \begin{enumerate}
   \item [2.1.] Follow Subsection 3.1 in
	   \citet{Hubert2012deterministic}  to obtain six robust, well-conditioned initial location
		 estimates $\mb^i$ and scatter estimates
		 $\mathbf{S}^i$ ($i=1,\ldots,6$). Use GLASSO to transform the scatter estimate into a precision matrix  $\mathbf{P}^i$ and denote the corresponding regularization parameter by $\lambda^i$.
   \item [2.2.] Compute for each subset, the extended BIC criterion and set $\lambda$ to the $\lambda^i$ of the subset with lowest extended BIC criterion.  
   \item [2.3.] For all initial subsets $H_0^i$, use $\lambda$ and
				repeat the generalized RMCD C-steps   
				until convergence. Denote the resulting subsets as $H^i\;$.
   \item [2.4.] Let $H_{RMCD}$ be the subset with largest GLASSO objective function (penalized log-likelihood)  among the candidate subsets.
   \end{enumerate}
\item [3.] From $H_{RMCD}$ compute the final RMCD estimates of location, scale and
precision as in \citet{croux2012}.
\end{enumerate}
\end{document}